\documentclass[prl,showpacs,twocolumn,amsmath,amssymb]{revtex4}
\usepackage[T1]{fontenc}
\usepackage[latin9]{inputenc}
\usepackage{booktabs}
\usepackage{amsmath}
\usepackage{graphicx}
\usepackage{amsfonts}
\usepackage{dcolumn}
\usepackage{bm}
\usepackage{float} 

\begin{document}

\title{Threshold of a Random Laser with Cold Atoms}

\author{Luis S. Froufe-P\'{e}rez$^1$, William Guerin$^2$, R\'{e}mi
Carminati$^3$ and Robin Kaiser$^{2}$}

\email{Robin.Kaiser@inln.cnrs.fr}
 \affiliation{\mbox{$^1$Instituto
de Ciencia de Materiales de Madrid,
CSIC, Sor Juana In\'{e}s de la Cruz~3, Cantoblanco, Madrid 28049, Spain,}\\
\mbox{$^2$Institut Non Lin\'{e}aire de Nice, CNRS and Universit\'{e} de
Nice Sophia-Antipolis, 1361 route des Lucioles,
06560 Valbonne, France,}\\
\mbox{$^3$Institut Langevin, ESPCI, CNRS UMR 7587, Laboratoire
d'Optique Physique, 10 rue Vauquelin, 75231 Paris Cedex 05,
France.}}

\date{\today}

\begin{abstract}
We address the problem of achieving an optical random laser with a
cloud of cold atoms, in which gain and scattering are provided by
the same atoms. The lasing threshold can be defined using the
on-resonance optical thickness $b_0$ as a single critical
parameter. We predict the threshold quantitatively, as well as
power and frequency of the emitted light, using two different
light transport models and the atomic polarizability of a
strongly-pumped two-level atom. We find a critical $b_0$ on the
order of 300, which is within reach of state-of-the-art cold-atom
experiments. Interestingly, we find that random lasing can already
occur in a regime of relatively low scattering.
\end{abstract}



\maketitle

Random lasing occurs when the optical feedback due to multiple
scattering in a gain medium is strong enough so that gain in the
sample volume overcomes losses through the surface. Since its
theoretical prediction by Letokhov \cite{Letokhov:1968}, great
efforts have been made to experimentally demonstrate this effect
in different kinds of systems
\cite{Markushev_1986,Lawandy_Nature_1994,Cao:1999,Wiersma:2001,Gottardo:2008},
as well as to understand the fundamentals of random lasing
\cite{wiersma_PRE_1996,Cao_PRL_2001}. The broad interest of this
topic is driven by potential applications (see
\cite{Wiersma_NatPhys_2008} and references therein) and by its
connections to the subject of Anderson localization
\cite{Conti:2008}. State-of-the-art random lasers
\cite{Wiersma_NatPhys_2008} are usually based on condensed matter
systems, and feedback is provided by a disordered scattering
medium, while gain is provided by an active material lying in the
host medium or inside the scatterers. In general, scattering and
gain are related to different physical entities.

Another system that can be considered for achieving random lasing
is a cold atomic vapor, using magneto-optical traps, where
radiation trapping \cite{Labeyrie:2003} as well as lasing
\cite{Hilico:1992,Guerin:2008} have been demonstrated. One
advantage is the ability to easily model the microscopic response
of the system components, which can be extremely valuable to fully
understand the physics of random lasers. However, in such system,
the ability to combine gain and multiple scattering at the same
time is not obvious, as both should be provided by the same atoms.
The purpose of this Letter is to address this issue
quantitatively. Note that even though new interesting features
appear when coherent feedback is involved \cite{Cao:2005}, we will
consider only incoherent (intensity) feedback.

Following Letokhov's theory, we consider a homogenous, disordered
and active medium of size $L$. The random lasing threshold is
governed by two characteristic lengths: the elastic scattering
mean free path $\ell_\mathrm{sc}$ \cite{Rossum:1999,footnote_ltr}
and the linear gain length $\ell_\mathrm{g}$ ($\ell_\mathrm{g} <0$
in the cases of absorption or inelastic scattering). In the
diffusive regime, defined as $L \gg \ell_\mathrm{sc}$, the lasing
threshold is reached when the unfolded path length,
$L^2/\ell_\mathrm{sc}$, becomes larger than the gain length. More
precisely, the threshold is given by \cite{Letokhov:1968,Cao:2003}
$L_\mathrm{eff} > \beta \pi \sqrt{\ell_\mathrm{sc}\,
\ell_\mathrm{g} /3}$,
where $\beta$ is a numerical factor that depends on the geometry
of the sample ($\beta=1$ for a slab, $\beta=2$ for a sphere), and
$L_\mathrm{eff} = \eta L$ is the effective length of the sample,
taking into account the extrapolation length \cite{Rossum:1999}.
Another important length scale is the extinction length, as
measured by the forward transmission of a beam through the sample,
$T = e^{-L/\ell_\mathrm{ex}}$. The extinction length is given by
$\ell_\mathrm{ex}^{-1} = \ell_ \mathrm{sc}^{-1}
-\ell_\mathrm{g}^{-1}$.

Let us consider now a homogeneous atomic vapor, constituted by
atoms of polarizability $\alpha(\omega)$ at density $\rho$,
submitted to a homogeneous pump field. The extinction length and
the scattering mean free path are related to their corresponding
cross-section $\sigma$ by $\ell_\mathrm{ex,sc}^{-1} = \rho\,
\sigma_\mathrm{ex,sc}$, with $\sigma_\mathrm{ex}(\omega) = k_0
\times \mathrm{Im}[\alpha(\omega)]$ and
$\sigma_\mathrm{sc}(\omega) = k_0^4/6\pi \times
|\alpha(\omega)|^2$ \cite{Lagendijk:1996}. As we consider only
quasi-resonant light, we use only the wave vector $k_0=\omega_0/c$
with $\omega_0$ the atomic frequency. We also define a
dimensionless atomic polarizability $\tilde{\alpha}$ such that
$\alpha = \tilde{\alpha} \times 6\pi/k_0^3$ (we omit the
dependence on $\omega$ in the following). As it is an intrinsic
parameter of the cloud, it is convenient to use the on-resonance
optical thickness $b_0 = \rho \sigma_0 L$, where $\sigma_0 =
6\pi/k_0^2$ is the resonant scattering cross-section (without pump
laser). Using these quantities, the threshold condition writes
\begin{equation}\label{eq.b0cr}
\eta b_0 > \frac{\beta \pi}{\sqrt{3 |\tilde{\alpha}|^2 \, \left(
|\tilde{\alpha}|^2-\mathrm{Im}(\tilde{\alpha}) \right)}} \, .
\end{equation}
Moreover, we have $L/\ell_\mathrm{sc} = b_0 |\tilde{\alpha}|^2$
and $\eta = 1 +  2 \xi / \left[L/\ell_\mathrm{sc} + 2 (\beta-1)
\xi \right]$ with $\xi \simeq 0.71$ for $L > \ell_\mathrm{sc}$
\cite{zweifel,R_eff}. Note that deeply in the diffusive regime ($L
\gg \ell_\mathrm{sc}$), $\eta \sim 1$.

Eq. (\ref{eq.b0cr}) is the first result of this Letter. It shows,
in the diffusive regime, the existence of a threshold of random
lasing as soon as the medium exhibits gain, i.e.,
$|\tilde{\alpha}|^2-\mathrm{Im}(\tilde{\alpha}) > 0$. This
threshold is given by a critical on-resonance optical thickness,
expressed as a function of the atomic polarizability only.
Interestingly, the condition $\mathrm{Im}(\tilde{\alpha}) < 0$,
corresponding to single-pass amplification ($T>1$), is not a
necessary condition.

The previous result is general and does not depend on a particular
pumping mechanism or atomic model. Let us now specify a gain model
that will allow numerical evaluations of the lasing threshold and
of the features of the emitted light. We shall use the simplest
case of strongly-pumped two-level atoms, for which the normalized
atomic polarizability at frequency $\omega$ can be written
analytically (assuming a weak ``probe'' intensity)
\cite{mollow_pra},
\begin{multline}\label{eq.mollow}
\tilde{\alpha}(\delta,\Delta,\Omega) = -\frac{1}{2}\,
\frac{1+4\Delta^2}{1+4\Delta^2+2\Omega^2}\\  \times
\frac{(\delta+i)(\delta-\Delta+i/2)-\Omega^2\delta/(2\Delta-i)}
{(\delta+i)(\delta-\Delta+i/2)(\delta+\Delta+i/2)-\Omega^2(\delta+i/2)}
\; . \raisetag{40pt}
\end{multline}
In this expression, $\Delta =
(\omega_\mathrm{p}-\omega_{0})/\Gamma$ is the normalized detuning
between the pump frequency $\omega_\mathrm{p}$ and the atomic
transition $\omega_0$ of linewidth $\Gamma$, $\delta =
(\omega-\omega_\mathrm{p})/\Gamma$ is the normalized detuning
between the considered ``probe'' frequency and the pump, and
$\Omega$ is the Rabi frequency, normalized by $\Gamma$, associated
with the pump-atom interaction. For a strong enough pumping power,
this atomic polarizability allows for single-pass gain, when
$\textrm{Im}(\tilde{\alpha})$<0. This gain mechanism is referred
as ``Mollow gain'' \cite{Guerin:2008,mollow_pra} and corresponds to
a three-photon transition (population inversion in the
dressed-state basis).

\begin{figure}[t] \includegraphics{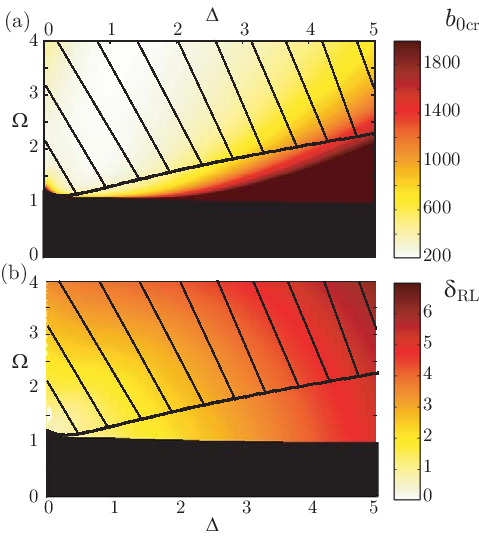}
\caption{Threshold of random lasing based on Mollow gain
\cite{mollow_pra}, calculated for each pair of pumping parameters
$\Delta$ (detuning) and $\Omega$ (Rabi frequency) with Eqs.
(\ref{eq.b0cr},\ref{eq.mollow}). Only the $\Delta >0$ part is
represented. (a) Critical optical thickness $b_{0\mathrm{cr}}$ to
allow lasing. (b) Detuning $\delta_\mathrm{RL}$ of the random
laser from the pump frequency. The black area corresponds to a
forbidden region (no gain). The hatched part corresponds to
parameters for which the diffusion approximation is \textit{a
priori} not reliable.}\label{fig.seuil_letokhov} \end{figure}

For each couple of pumping parameters $\{\Delta, \Omega \}$, the
use of the polarizability (\ref{eq.mollow}) into the threshold
condition (\ref{eq.b0cr}) allows the calculation of the critical
on-resonance optical thickness $b_0$ as a function of $\delta$.
Then, the minimum of $b_0$ and the corresponding $\delta$
determine the optical thickness $b_{0\mathrm{cr}}$ that the cloud
must overcome to allow lasing, and the frequency
$\delta_\mathrm{RL}$ of the random laser at threshold. The result
is presented in Fig. \ref{fig.seuil_letokhov} for a spherical
geometry ($\beta = 2$). The result for $b_{0\mathrm{cr}}$ is
independent of the sign of $\Delta$ and we only show the region
$\Delta>0$. The minimum optical thickness that allows lasing is
found to be $b_{0\mathrm{cr}} \approx 200$ and is obtained for a
large range of parameters, approximately along the line $\Omega
\approx 3 \Delta$. The optimum laser-pump detuning is near the
gain line of the transmission spectrum, i.e., $\delta_\mathrm{RL}
\sim \mathrm{sign}(\Delta) \sqrt{\Delta^2+\Omega^2}$ (a small
shift compared to the maximum gain condition is due to the
additional constraint of combined gain and scattering).

The obtained critical optical thickness is achievable with current
technology \cite{Barrett:2001}, showing that random lasing is
possible in a system of cold atoms with Mollow gain. As this
result has been obtained using the diffusion approximation, the
condition $L/\ell_\mathrm{sc} = b_0 |\tilde{\alpha}|^2 \gg 1$ must
be satisfied. This is not the case in the full range of random
lasing parameters that we have found. For example, with $\Delta
\approx 1$ and $\Omega \approx 3$, the critical optical thickness
is almost minimum, $b_{0cr} = 213$, but $L/\ell_\mathrm{sc}
\approx 0.44$. In this case, the threshold defined by
Eq.~(\ref{eq.b0cr}) is at best unjustified, at worst wrong. In
order to identify in Fig. \ref{fig.seuil_letokhov} the region in
which the approach should be valid, we have hatched the area
corresponding to $L/\ell_\mathrm{sc} < 3$. Note that random lasing
is still expected in this region, but for a larger on-resonance
optical thickness, that would allow the fulfillment of the
diffusive condition. The minimum optical thickness in the region
of parameters compatible {\it a priori} with the diffusion
approximation is 347, and is located in the vicinity of $\{\Delta
= 1, \Omega = 1.2 \}$.

This first evaluation demonstrates the need for a more refined
transport model. In the following, we use the approach introduced
in Ref. \cite{Remi_PRA_2007}, that is based on the radiative
transfer equation (RTE). The RTE is a Boltzmann-type transport
equation \cite{Chandrasekhar}, that has a larger range of validity
with respect to the ratio $L/\ell_\mathrm{sc}$ than the diffusion
equation~\cite{Remi_JOSA_2004}.

Letokhov's diffusive theory \cite{Letokhov:1968,Cao:2003} and the
RTE-based theory \cite{Remi_PRA_2007} of random lasing both rely
on a modal expansion of the solution of the transport equation. In
order to compare the predictions of both models, we focus on the
slab geometry ($\beta=1$) since the modal expansion of the RTE is
well known in this case~\cite{zweifel} (to our knowledge, no
simple expansion is available for a sphere in the RTE approach).
The modal approach consists in looking for solutions of the form
$\Psi_s(z,{\bf u},t)= \phi_{\kappa,s}({\bf u}) \, \exp(i\kappa z)
\, \exp(st)$, where $\Psi(z,{\bf u},t)$ is the specific intensity
($z$ is the distance from the slab surface and ${\bf u}$ denotes a
propagation direction). For a given real $\kappa$, $s(\kappa)$ and
$\phi_{\kappa,s}$ form a set of eigenvalues and eigenfunctions of
the RTE. If one denotes by $s_0(\kappa)$ the eigenvalue
corresponding to the mode with the longest lifetime in the passive
system, a laser instability appears when $s_0(\kappa)
>0$ in the presence of gain. The lasing threshold is defined by
the condition $s_0(\kappa)=0$. For isotropic scattering, this
eigenvalue has an analytical expression valid for $\kappa\,
\ell_\mathrm{sc} < \pi/2$~\cite{zweifel,Remi_PRA_2007}:
\begin{equation}\label{eq.s0RTE}
s_0(\kappa)/c = \ell_\mathrm{g}^{-1} - \left[
\ell_\mathrm{sc}^{-1} - \kappa/\tan(\kappa\,\ell_\mathrm{sc})
\right]
\end{equation}
where $c$ is the energy velocity. For a slab of width $L$, the
dominant mode corresponds to $\kappa \! =\! \pi/L_\mathrm{eff}\!
=\! \pi/(L +2\xi \ell_\mathrm{sc})$. In practice, this
determination of $\kappa$ is meaningful as long as $\xi=0.71$ can
be taken as a constant (independent on $L$), which is the case for
$L>\ell_\mathrm{sc}$. This condition sets the limit of accuracy of
the modal RTE approach.

The diffusive result is recovered from the RTE approach in the
limit $\kappa \ell_\mathrm{sc} \ll 1$~\cite{Remi_JOSA_2004}. A
first order expansion of Eq.~(\ref{eq.s0RTE}) yields
$s_0^\mathrm{(DA)}(\kappa)/c =\ell_\mathrm{g}^{-1} -\kappa^2
\ell_\mathrm{sc}/3$,
where the superscript (DA) stands for diffusion approximation. The
condition $s_0^\mathrm{(DA)}(\kappa=\pi/L_\mathrm{eff}) = 0$ leads
to Letokhov's threshold, with $\beta=1$.

The comparison between the RTE and diffusive approaches deserves
two comments. Firstly, the gain contribution to $s_0(\kappa)$
(first term in Eq. (\ref{eq.s0RTE})) is the same in both models.
Secondly, the scattering contribution (second term in Eq.
(\ref{eq.s0RTE})) is larger in the RTE model by a factor of at
most $1.13$ (when $L \sim \ell_\mathrm{sc}$). Thus, the correction
introduced by the RTE model, compared to the diffusion
approximation, is relatively small, as it corresponds to an
increase of $\eta \, b_{0\mathrm{cr}}$ of at most a few percents.
This means that the diffusive model gives accurate results down to
$L \sim \ell_\mathrm{sc}$, and that in cold-atom systems, random
lasing can occur even in a regime of low scattering.

\begin{figure}[t]
\includegraphics{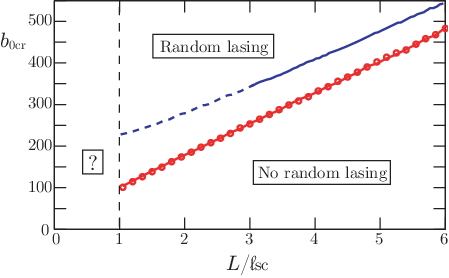}
\caption{Critical optical thickness for different geometries and
transport models. The lower, red curve corresponds to the slab
geometry (width $L$), with RTE model (continuous line) and
diffusive model (open circle). The upper, blue curve corresponds
to the sphere geometry (diameter $L$), with the diffusive model.
The part with $L/\ell_\mathrm{sc}<3$ is dashed, as the model may
not be reliable.} \label {fig.b0_comparison}
\end{figure}

In Fig. \ref{fig.b0_comparison}, we compare the minimum optical
thickness obtained with both models for the slab geometry and with
the diffusive model for the sphere geometry. To put forward the
domain of validity in each case, we plot the results as a function
of $L/\ell_\mathrm{sc}$. As expected, in the range
$L>\ell_\mathrm{sc}$ the threshold predicted by the RTE for the
slab geometry is only slightly larger than the one given by the
diffusion approximation, so that the two curves can hardly be
distinguished. For the sphere geometry, we dashed the part
corresponding to the domain where the diffusive model is \textit{a
priori} not reliable, i.e., $L/\ell_\mathrm{sc} < 3$.
Nevertheless, by generalizing the conclusion obtained with the
slab geometry, we reasonably expect the threshold to be located
between 250 and 300.

Let us now turn to a first characterization of such a random
laser. An important quantity to be investigated is the emitted
power as a function of the pumping power. In the stationary regime
(continuous pumping) we numerically solve the optical Bloch
equations for a strongly-pumped two-level atom (without using the
weak ``probe'' approximation that leads to Eq. (\ref{eq.mollow}))
to obtain the polarizability at the lasing frequency, including
the gain saturation induced by the random laser intensity. Above
threshold, the laser intensity \emph{in the medium}
$I_\mathrm{RL}^\mathrm{(in)}\propto |\Omega_\mathrm{RL}|^2$ is
determined by the condition $s_0(\kappa,
|\Omega_\mathrm{RL}|^2)=0$ ($s_0$ would be positive without gain
saturation). The obtained intensity is analogous to the
intra-cavity intensity of a standard laser, and thus does not
correspond to the emitted power $P_\mathrm{RL}^\mathrm{(out)}$. At
equilibrium, gain compensates losses, and
$P_\mathrm{RL}^\mathrm{(out)}$ is equal to the generated power,
related to the gain cross-section $\sigma_\mathrm{g}$, i.e.,
$P_\mathrm{RL}^\mathrm{(out)} \propto \sigma_\mathrm{g}
|\Omega_\mathrm{RL}|^2$ with $\sigma_\mathrm{g} = \sigma_0
\left(|\tilde{\alpha}|^2-\mathrm{Im}(\tilde{\alpha}) \right)$.

\begin{figure}[t]
\includegraphics{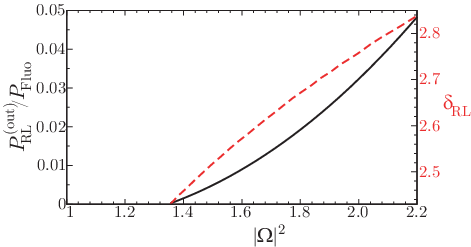}
\caption{Continuous line: Emitted random laser power normalized to the pump fluorescence
power, as a function of the pump intensity. Dashed line: Normalized laser detuning
$\delta_\mathrm{RL}$. The random medium is a spherical cloud of
two level atoms with an on-resonance optical-thickness
$b_{0}=650$.}\label{fig:3}
\end{figure}

In order to know if the laser signal can be extracted from the
background fluorescence, it is particularly relevant to compare
the emitted laser power with the pump-induced fluorescence
$P_\mathrm{Fluo} \propto \sigma_0
|\Omega|^2/(1+4\Delta^2+2|\Omega|^2)$. From this, we compute the
ratio
\begin{equation}\label{eq.intensite}
\frac{P_\mathrm{RL}^\mathrm{(out)}}{P_\mathrm{Fluo}}=
\frac{|\Omega_\mathrm{RL}|^2}{|\Omega|^2}
\left(|\tilde{\alpha}|^{2}-\textrm{Im}(\tilde{\alpha})\right)
\left(1+4\Delta^2+2|\Omega|^2\right) \; .
\end{equation}
We plot the result in Fig. \ref{fig:3} as a function of
$|\Omega|^2$, for a pump detuning $\Delta=1$. To obtain Eq.
(\ref{eq.intensite}), we assume that both pump and laser
intensities are homogeneously distributed across the whole system.
We also consider only the optimum random laser frequency, thus
neglecting the spectral width of the random laser or any
interaction between different random laser frequencies. Hence we
neglect several effects as mode competition
\cite{Stone_Science_2008} and inelastic scattering of the laser
light. Nevertheless we think that the order or magnitude of the
ratio laser-to-fluorescence powers can be realistic for actual
experiments, at least as long as only one mode of the laser is
active \cite{Stone_Science_2008}. For the chosen set of
parameters, this ratio is more than 5\% and hence laser emission
should be measurable. Its distinction from the pump-induced
fluorescence can be made by looking at the spectrum of the emitted
light. Another interesting prediction of this model is that the
laser emission frequency shifts as the pump intensity is increased
[Fig. \ref{fig:3}]. This corresponds to the shift of the maximum
gain of the Mollow polarizability.

In summary, we have established the possibility of achieving
random lasing with cold atoms. The random laser threshold is
described by a single critical parameter, the on-resonance optical
thickness $b_0$. In the particular case of a gain mechanism based
on a strongly-pumped two-level atom (Mollow gain), our model
predicts a critical $b_0\sim 300$. Such an optical thickness is
achievable in current cold-atoms experiments, \textit{e.g.} by
using crossed dipole traps \cite{Barrett:2001}. We have also
determined the basic features of the emitted light above
threshold, showing that the random laser emission should be
measurable.

Another interesting result is that, due to the large gain, lasing
can be obtained with a low feedback (low amount of scattering,
i.e., $L \sim \ell_\mathrm{sc}$). This regime is similar to that
encountered in certain semiconductor lasers with a very poor
cavity, and is different from the working regime of random lasers
realized to date. This new regime could be numerically
investigated by RTE-based simulations \cite{Remi_JOSA_2004}.

Finally, let us stress that the model developed here has several
limitations, so that the numbers should be considered as
first-order estimates. Firstly, we have considered monochromatic
pumping, thus neglecting inelastic scattering from the pump. The
inelastically-scattered photons may have a non-negligible
influence on the atomic response, as shown in
\cite{Davidson:1999}. Secondly, the RTE model needs to be extended
to a sphere geometry, and to a medium with inhomogeneous density
and/or pumping. This would require a full numerical solution of
coupled RTEs for the pump and probe beams \cite{Remi_PRA_2007}. We
also outline that the case of the Mollow gain was chosen for the
sake of simplicity, whereas other gain mechanisms might be more
adapted for the search of random lasing, such as Raman gain or
parametric gain \cite{Guerin:2008}. Each gain mechanism has its
advantages and drawbacks, but the degrees of freedom they offer,
together with the first estimates presented here, make us
confident that the experimental realization of a cold-atom random
laser is possible with current technology.

The authors thank S. Skipetrov, J. J. S\'{a}enz, R. Pierrat and F. Michaud for
useful discussions. L.S.F. acknowledges the financial support of
Spanish ministry of science and innovation through its Juan de la
Cierva program. This work is supported by ANR-06-BLAN-0096.

\end{document}